\documentstyle[12pt]{article}

   \font\tenmsb=msbm10 scaled\magstep 1
   \font\sevenmsb=msbm7 scaled \magstep 1
   \font\faivemsb=msbm5 scaled \magstep 1
\newfam\msbfam
      \textfont\msbfam=\tenmsb
      \scriptfont\msbfam=\sevenmsb
      \scriptscriptfont\msbfam=\faivemsb
\def\Bbb#1{{\fam\msbfam #1}}
\font\tengothic=eufm10 scaled\magstep 1
\font\sevengothic=eufm7 scaled\magstep 1
\newfam\gothicfam
      \textfont\gothicfam=\tengothic
      \scriptfont\gothicfam=\sevengothic

\newcommand{\be}{\begin{equation}}
\newcommand{\ee}{\end{equation}}
\newcommand{\vp}{\varphi}
\newcommand{\ep}{\varepsilon}
\newcommand{\ra}{\rightarrow}
\newcommand{\ox}{\overline x}
\newcommand{\dgr}{\dagger}
\newcommand{\om}{\omega}
\newcommand{\gm}{\gamma}
\newcommand{\al}{\alpha}

\begin{document}

\begin{center}
{\Large{\bf Evolutional Entanglement in Nonequilibrium Processes} \\ [5mm]

V.I. Yukalov} \\ [3mm]

{\it Bogolubov Laboratory of Theoretical Physics \\

Joint Institute for Nuclear Research, Dubna 141980, Russia}

\end{center}

\vskip 1cm

\begin{abstract}

Entanglement in nonequilibrium systems is considered. A general definition
for entanglement measure is introduced, which can be applied for
characterizing the level of entanglement produced by arbitrary operators.
Applying this definition to reduced density matrices makes it possible to
measure the entanglement in nonequilibrium as well as in equilibrium
statistical systems. An example of a multimode Bose-Einstein condensate
is discussed.

\end{abstract}

\section{Introduction}

The concept of entanglement [1] is believed to play an important role in
quantum information processing and quantum computing, because of which much
efforts has been devoted to quantifying entanglement [2--8]. The latter is
usually measured by a kind of reduced or relative entropy and is considered
for bipartite systems [9,10]. The choice of entropy as a measure of
entanglement looks reasonable since entropy characterizes the complexity
of both equilibrium and nonequilibrium systems[11]. However, measuring
entanglement by means of a reduced or relative entropy is relatively
straightforward only for rather simple bipartite systems. Moreover, there
does not exist a measure of many-body entanglement for multipartite systems.
It is even less clear how to quantify entanglement in nonequilibrium
statistical systems.

The aim of the present communication is to introduce a general entanglement
measure that would be valid in arbitrary cases and to illustrate its
application to a nonequilibrium system. As an example of the latter,
a multimode Bose-Einstein condensate of trapped atoms is considered.

\section{Entanglement Measure}

Before going to physical applications, it is useful to define what actually
entanglement means from the general mathematical point of view. As far as one
usually speaks about the entanglement of some states, it is necessary, first,
to specify what states are to be entangled.

Consider a set of objects, called parts, which are enumerated by an index
$i=1,2,\ldots,p$. Each part is characterized by a Hilbert {\it space of
single-partite quantum states}
\be
\label{1}
{\cal H}_i \equiv \overline{\cal L}\{\; |n_i>\} \; ,
\ee
being a closed linear envelope of a single-partite basis $\{\;|n_i>\}$ of
quantum states $|n_i>$. The {\it space of composite-system quantum states}
is a subspace
\be
\label{2}
{\cal H} \subset {\cal H}^p \equiv \otimes_{i=1}^p {\cal H}_i
\ee
of the $p$-fold tensor product ${\cal H}^p$ of spaces (1). In particular,
${\cal H}$ may coincide with ${\cal H}^p$. But ${\cal H}$ becomes a subspace
of ${\cal H}^p$ if the tensor-product space is complimented with some
selection rules restricting admissible states. For example, such selection
rules may consist of the requirement of specific symmetry properties.
A subspace of a tensor-product space can be called {\it incomplete
tensor-product space} [12].

From the composite-system space ${\cal H}$, we separate out the set
${\cal D}\subset{\cal H}$ of disentangled states, thus, obtaining the {\it
disentangled set}
\be
\label{3}
{\cal D} \equiv \left\{ \otimes_{i=1}^p \; \vp_i\; | \;
\forall \vp_i\in {\cal H}_i\right \} \; .
\ee
The states of ${\cal D}$ have the structure of tensor products of
$\vp_i\in{\cal H}_i$, but ${\cal D}$ does not include linear combinations of
such products. This principally distinguishes the disentangled set ${\cal D}$
from the total  space ${\cal H}$. The latter, in addition to the tensor
products of quantum states, contains as well their linear combinations. The
compliment ${\cal H}\setminus {\cal D}$ composes the set of entangled states.

The meaning of the word entanglement is that the disentangled states from
${\cal D}$ are, by means of some transformation, converted into the
entangled states of ${\cal H}\setminus {\cal D}$. Transformations are
accomplished by operators. Hence, mathematically, entanglement implies that
an operator $A$ acting on ${\cal D}$ transforms it into ${\cal H}\setminus
{\cal D}$. Thus, the concept of entanglement, to be mathematically correct,
must include the definition of disentangled states, to be entangled, and
the specification of an operator, producing this entanglement.

Let a bounded operator $A$ be defined on ${\cal H}$. Because it is bounded, it
possesses a finite norm
$$
||A||_{{\cal H}}\; \equiv \sup_{||\vp||_{{\cal H}}=1} ||A\vp||_{{\cal H}} \; ,
$$
where $||\vp||_{{\cal H}}$ is a vector norm of $\vp$ on ${\cal H}$. In
addition to the norm $A$ on ${\cal H}$, we may define the norm of $A$ on
${\cal D}$, that is,
$$
||A||_{\cal D} \; \equiv \sup_{||f||_{\cal D}=1}\; ||Af||_{\cal D} \; .
$$

It looks that to quantity the entanglement caused by an operator, we shall
need to invoke the operator norms. This understanding comes from the fact
that the operator norms are involved in measuring the order associated with
operators, which is characterized by the {\it operator order indices} [13].
These indices can be introduced for arbitrary operators. In particular, they
can be defined for density matrices, thus, describing ordering in physical
systems [14].

The entanglement, produced by an operator, should be described by comparing
the performance of this operator on the disentangled set with the action of
a related operator that does not entangle the states of ${\cal D}$.
A nonentangling operator should have the structure of the tensor product
of operators acting on ${\cal H}_i$. To this end, we construct the {\it
product operator}
\be
\label{4}
A^{\otimes} \; \equiv
\frac{{\rm Tr}_{\cal H}\; A}{{\rm Tr}_{\cal D}\;\otimes_{i=1}^p A_1^i} \;
\otimes_{i=1}^p\; A_1^i \; ,
\ee
in which
$$
A_1^i \equiv const \; {\rm Tr}_{ \{ {\cal H}_{j\neq i} \} }\; A
$$
and the form (4) is chosen so that to satisfy the normalization condition
$$
{\rm Tr}_{{\cal H}} \; A = {\rm Tr}_{\cal D} \; A^\otimes \; .
$$
Comparing the action on ${\cal D}$ of an operator $A$ with that of its
nonentangling counterpart (4), we define the {\it entanglement measure}
\be
\label{5}
\ep(A) \equiv \log \; \frac{||A||_{\cal D}}{||A^\otimes||_{\cal D}} \; ,
\ee
where the logarithm can be taken with respect to any base, say, to the base 2.

It is straightforward to show that the entanglement measure (5) possesses
the properties that are natural  for being such a measure. First of all, this
measure is {\it semipositive},
\be
\label{6}
\ep(A)\geq 0 \; .
\ee
Second, the measure is {\it continuous} in the sense that if for any operator
$A$ on ${\cal H}$ there exists a family $\{ A(t)\}$ of operators $A(t)$,
parametrized with $t\in\Bbb{R}$, so that
$$
||A(t)||_{\cal D} \; \ra ||A||_{\cal D} \qquad (t\ra 0) \; ,
$$
then
\be
\label{7}
\ep(A(t)) \ra \ep(A) \qquad (t\ra 0) \; .
\ee
Third, a {\it nonentangling} operator, having the structure of a tensor
product $A^\otimes$, does not produce entanglement,
\be
\label{8}
\ep\left ( A^\otimes \right ) = 0 \; .
\ee
The latter property may ne generalized to the case when
$A=\oplus_\nu p_\nu A^\otimes_\nu$ is a linear combination of the operators
$A_\nu^\otimes$ such that
$$
|| A_\nu^\otimes ||_{\cal D} \; = ||A^\otimes||_{\cal D} \; , \qquad
\sum_\nu \; |p_\nu| = 1 \; .
$$
In that case, one has
$$
\ep\left ( \oplus_\nu \; p_\nu A_\nu^\otimes \right ) = 0 \; .
$$
Fourth, the entanglement is {\it additive}, which means the following. Let
$A=\otimes_\nu A_\nu$, then
\be
\label{9}
\ep \left ( \otimes_\nu A_\nu \right ) = \sum_\nu \; \ep(A_\nu) \; .
\ee
Finally, the measure is {\it invariant} under local unitary operations
$U_i$, such that $U_i^+ U_i =1$, for which we have
\be
\label{10}
\ep \left ( \otimes_{i=1}^p U_i^+ A \otimes_{i=1}^p U_i \right ) =
\ep(A) \; .
\ee

In this way, the entanglement measure (5) can be defined for an arbitrary
operator. In physical applications, one may consider the entanglement produced
by any operator from the algebra of observables, for instance by a Hamiltonian,
by a spin operator, and so on. One may also investigate the entanglement caused
by statistical operators and by density matrices.

As an illustration, we may consider a simple case of a bipartite system, when
${\cal H}={\cal H}_1\otimes{\cal H}_2$. Let each ${\cal H}_i$ be a separable
Hilbert space of dimension $d_i$, which is a span of an orthonormal basis
$\{\; |\vp_n^i>\}$. Consider entanglement realized by a von Neumann statistical
operator
$$
\hat\rho_{BP} = |BP><BP| \; , \qquad |BP>\; \in {\cal H}\; ,
$$
describing a pure statistical state of the bipartite system. The wave function
of a bipartite system can be presented (see e.g. [14]), involving the Schmidt
decomposition, as the biorthogonal sum
$$
|BP> \; = \sum_{n=1}^d \; c_n\; |\vp_n^1>\; \otimes\; |\vp_n^2> \; ,
$$
in which $d\equiv \min_i d_i$ and $\sum_{n=1}^d|c_n|^2=1$. Following the
procedure, described above, we have
$$
\hat\rho_1^i \equiv {\rm Tr}_{{\cal H}_{j\neq i}} \; \hat\rho_{BP} =
\sum_{n=1}^d \; |c_n|^2\; \left | \vp_n^i><\vp_n^i\right | \; .
$$
The corresponding norms are
$$
||\hat\rho_{BP}||_{\cal D} \; = ||\hat\rho_1^i||_{{\cal H}_i} \; =
\sup_n |c_n|^2 \; . $$
Then for the entanglement measure (5), we get
$$
\ep\left (\hat\rho_{BP}\right ) = - \log\sup_n |c_n|^2 \; .
$$
This varies in the interval
$$
0\leq \ep\left (\hat\rho_{BP}\right ) \leq \log d \; .
$$
Maximal entanglement occurs when $|c_n|^2=1/d$. One usually quantifies
entanglement in a pure bipartite system by the reduced von Neumann entropy
$$
S_N^i \equiv -{\rm Tr}_{{\cal H}_i}\; \hat\rho_1^i\; \log\hat\rho_1^i =
- \sum_{n=1}^d |c_n|^2\; \log|c_n|^2 \; .
$$
The latter coincides with measure (5) for the maximally entangled state, when
$S_N^i=\ep(\hat\rho_{BP})=\log d$. Note that for a mixed bipartite state the
reduced entropy $S_N^i$ is, generally, different for $i=1$ and $i=2$, thus,
becoming not well defined as an entanglement measure, while
$\ep(\hat\rho)$ is always well defined.

\section{Evolutional Entanglement}

In general, if the considered operator $A=A(t)$ depends on time $t$, the
entanglement measure (5) will be a function of time $\ep(A(t))$, displaying
the temporal evolution of entanglement. To study the evolutional entanglement
for physical systems, we may employ the reduced density matrices.

Consider a set $x^p\equiv\{ x_1,x_2,\ldots,x_p\}$ of variables characterizing
a physical system. Each $x_i\in{\cal X}_i$ pertains to a characteristic space
${\cal X}_i$. A $p$-order reduced density matrix
\be
\label{11}
\rho_p(t) =\left [ \rho_p\left ( x^p,\ox^p,t\right )
\right ]
\ee
is a matrix with respect to the variables $x^p$ and $\ox^p$, The matrix
elements being
\be
\label{12}
\rho_p\left ( x^p,\ox^p,t\right ) \equiv {\rm Tr}_{\cal F}\; \psi(x_1)\ldots
\psi(x_p)\hat\rho(t) \psi^\dgr(\ox^p)\ldots \psi^\dgr(\ox_1) \; ,
\ee
where the trace is over the Fock space, $\psi(x)$ is a field operator, and
$\hat\rho(t)$ is a statistical operator. The first-order density matrix is
\be
\label{13}
\rho_1^i(t) =\left [\rho_1\left (x_i,\ox_i,t\right ) \right ] \; ,
\ee
with the elements
\be
\label{14}
\rho_1\left ( x,\ox,t\right ) \equiv {\rm Tr}_{\cal F} \;
\psi(x) \hat\rho(t) \psi^\dgr(\ox) \equiv \; < \psi^\dgr(\ox)\; \psi(x) > \; .
\ee
One often calls the matrices (11) and (13) the $p$-particle and single-particle
density matrices. This in no way means that some concrete particles are
compulsory separated out of the system, but just designates the order of the
matrices. We may keep in mind a system of $N$ indistinguishable particles,
with the same characteristic space ${\cal X}$. Then a $p$-order density
matrix describes correlations between any $p$ particles from the ensemble
of $N$ identical particles.

The trace of $\rho_1^i(t)$ over the single-partite space (1) is
\be
\label{15}
{\rm Tr}_{{\cal H}_i}\; \rho_1^i(t) \equiv \sum_{n_i}\; < n_i\; |\rho_1^i(t) |
\; n_i>\; = \int \rho_1(x_i,x_i,t)\; dx_i = N \; .
\ee
Similarly, the trace of $\rho_p(t)$ over ${\cal H}$ is
\be
\label{16}
{\rm Tr}_{\cal H}\; \rho_p(t) = \int \rho_p\left ( x^p,x^p,t\right ) \; dx^p =
\frac{N!}{(N-p)!} \; .
\ee
The relation between the single-partite and $p$-partite density matrices reads
\be
\label{17}
\rho_1^i(t) = \frac{(N-p)!}{(N-1)!} \; {\rm Tr}_{\{ {\cal H}_{j\neq i}\} } \;
\rho_p(t) \; .
\ee
The product operator (4) becomes
\be
\label{18}
\rho^\otimes_p(t) = \frac{N!}{(N-p)!\; N^p} \; \otimes_{i=1}^p \;
\rho_1^i(t) \; .
\ee
Calculating the norms of density matrices, over the disentangled set (3), we
may define the single-partite basis $\{ |n_i>\}$ as formed by the eigenvectors
of $\rho_1^i(t)$. Then the latter can be presented as the diagonal expansion
\be
\label{19}
\rho_1^i(t) = \sum_{n_i} D_{n_i}(t) \; |n_i><n_i| \; .
\ee
For the entanglement measure (5), we have
\be
\label{20}
\ep(\rho_p(t)) = \log\;
\frac{(N-p)!N^p||\rho_p(t)||_{\cal D}}{N!\;\prod_{i=1}^p
||\rho_1^i(t)||_{{\cal H}_i}}\; .
\ee
This describes the level of entanglement between any $p$ parts from the system
of $N$ parts.

\section{Multimode States}

To illustrate the calculation of the entanglement measure (20), let us consider
the case of a multimode coherent system whose $p$-partite density matrix can be
reduced to the form
\be
\label{21}
\rho_p(t) = \frac{N!}{(N-p)!}\; \sum_n  w_n(t) |n\ldots n><n\ldots n| \; ,
\ee
in which the fractional mode populations $w_n(t)$ satisfy the properties
\be
\label{22}
0\leq w_n(t) \leq 1 \; , \qquad \sum_n w_n(t) = 1 \; .
\ee
As a physical application, we may keep in mind a system of $N$ trapped atoms
in Bose-Einstein condensate (see reviews [15--17]). At low temperature, when
practically all $N$ atoms are condensed, the system is in a coherent state
[17,18]. Several modes in such a system can be created in different ways,
e.g., by localizing atomic clouds in a multiwell potential formed by an
optical lattice [19,20], by mixing falling and reflected atomic wave packets
[21], and by other means. A controllable method of exciting topological
coherent modes in a trapped Bose-Einstein condensate is by modulating the
trapping potential with the help of resonant alternating fields [22--24].

For the density matrix (21), we have
\be
\label{23}
||\rho_p(t)||_{\cal D} = \frac{N!}{(N-p)!}\; \sup_n w_n(t) \; ,
\ee
and for the product matrix (18), we find
\be
\label{24}
||\rho_p^\otimes(t)||_{\cal D} = \frac{N!}{(N-p)!} \; \sup_n \; w_n^p(t) \; .
\ee
Therefore, the entanglement measure (20) becomes
\be
\label{25}
\ep(\rho_p(t)) = (1-p)\; \log\; \sup_n w_n(t) \; .
\ee
If the number of modes is $m\equiv\sum_n 1$, then the maximal entanglement
happens when all $w_n(t)$ are equal with each other, that is, are equal to
$1/m$. The entanglement measure (25) can take the values in the interval
\be
\label{26}
0 \leq \ep(\rho_p(t)) \leq (p-1)\; log\; m \; .
\ee
The temporal behaviour of measure (25) is prescribed by the evolution of the
fractional mode populations $w_n(t)$.

In order to be absolutely concrete, let us present an example explicitly
demonstrating the evolution equations for $w_n(t)$. For this purpose, let us
consider the resonant generation of {\it topological coherent modes} in a
trapped Bose-Einstein condensate [22--24]. An ensemble of coherent atoms is
described by the Gross-Pitaevskii equation [16--18]. The stationary solutions
to the latter define the topological coherent modes with a discrete spectrum of
energies $E_n$. The condensate is subject to the action of external alternating
fields with the frequencies tuned close to some of the transition frequencies
$\om_{mn}\equiv (E_m-E_n)/\hbar$. Suppose, for concreteness, that there are
two  resonant fields whose frequencies are tuned to the transition frequencies
$\om_{21}$ and $\om_{32}$, with the detunings $\Delta_{21}$ and $\Delta_{32}$,
respectively. In the resonance approximation, the Gross-Pitaevskii equation
can be reduced [22--24] to the system of equations for the mode amplitudes
$c_n(t)$ which define the fractional mode populations
\be
\label{27}
w_n(t) \equiv |c_n|^2 \; .
\ee
In the considered case of two resonant fields, connecting three coherent modes,
we have three mode populations, $w_1(t)$, $w_2(t)$, and $w_3(t)$. The evolution
equations for the latter can be obtained from the equations for the mode
amplitudes $c_n(t)$. The equations for the mode populations involve the ladder
variables
$$
h_1 \equiv 2c_1^*\; c_2\; \exp\{ i(\Delta_{21} t + \gm_{12} ) \} \; , \qquad
h_2 \equiv 2c_2^*\; c_3\; \exp\{ i(\Delta_{32} t + \gm_{23} ) \} \; ,
$$
\be
\label{28}
h_3 \equiv 2c_1^*\; c_3\; \exp\{ i(\Delta_{32} +\Delta_{21}) t +
i( \gm_{12}+\gm_{23}) \} \; ,
\ee
in which $\gm_{mn}$ are initial phases. For the mode populations (27), we
derive the evolution equations
$$
\frac{dw_1}{dt} = \frac{i}{4}\; b_{12}\left ( h_1^* - h_1 \right ) \; ,
$$
$$
\frac{dw_2}{dt} = \frac{i}{4}\; b_{23} \left ( h_2^* - h_2\right ) -\;
\frac{i}{4}\; b_{12}\; \left ( h_1^* - h_1 \right ) \; ,
$$
\be
\label{29}
\frac{dw_3}{dt} = -\; \frac{i}{4}\; b_{23} \;
\left ( h_2^* - h_2 \right )  \; ,
\ee
where $b_{mn}$ are the transition amplitudes due to external alternating
resonant fields. And the ladder variables (28) satisfy the equations
$$
i\; \frac{dh_1}{dt} = - h_1\; [ \al_{12} w_2 - \al_{21} w_1 +
( \al_{13} - \al_{23} )\; w_3 + \Delta_{21} ] - b_{12}\; ( w_2 - w_1 ) +
\frac{1}{2}\; b_{23}\; h_3 \; ,
$$
$$
i\; \frac{dh_2}{dt} = - h_2\; [ \al_{23} w_3 - \al_{32} w_2 +
( \al_{21} - \al_{31} )\; w_1 + \Delta_{32} ] - b_{23}\; ( w_3 - w_2 ) -\;
\frac{1}{2}\; b_{12}\; h_3 \; ,
$$
\be
\label{30}
i\; \frac{dh_3}{dt} = - h_3\; [ \al_{13} w_3 - \al_{31} w_1 +
( \al_{12} - \al_{32} )\; w_2 + \Delta_{32} + \Delta_{21} ] - \;
\frac{1}{2}\; b_{12}\; h_2 + \frac{1}{2}\; b_{23}\; h_1 \; ,
\ee
where $\al_{mn}$ are the transition amplitudes corresponding to the interatomic
interactions. The system of equations (29) and (30) contains nine real-valued
equations. However, by definitions (27) and (28), not all variables $w_n(t)$
and $h_n(t)$ are independent. Thus, we have the following relations
$$
|h_1|^2 = 4 w_1 \; w_2 \; , \qquad |h_2|^2 = 4w_2\; w_3\; , \qquad
|h_3|^2 = 4 w_3\; w_1 \; ,
$$
$$
h_1\; h_2 = 2w_2\; h_3\; , \qquad w_1 + w_2 + w_3 = 1 \; .
$$
As a result, Eqs. (29) and (30) can be reduced to an effective four-dimensional
dynamical system. Solving these equations, we can find the temporal behaviour
of the entanglement measure (25).

The solution of the system of nonlinear equations (29) and (30) requires
numerical calculations. It is not trivial even for a simpler two-mode case
[22--24], displaying such interesting effects as mode locking [22,24] and
critical dynamics [23--25]. The situation is essentially more complicated
for the three-mode case, considered here. In the frame of a brief
communication it is impossible to discuss the details of the temporal
properties of the mode populations (27), which will be done in a separate
publication. We may only mention that the behaviour of $w_n(t)$ is relatively
simple when $|b_{mn}|\ll 1$, being a type of Rabi oscillations. Respectively,
the entanglement measure (25) will oscillate with time. But with increasing
$|b_{mn}|$, the temporal evolution of this measure becomes much more
intricate.

\section{Conclusion}

A general definition for entanglement measure is advanced, which is valid for
quantifying entanglement realized by an arbitrary operator. We concentrate here
on discussing the case of nonequilibrium statistical systems, for which it is
reasonable to study the entanglement caused by reduced density matrices. As
an illustration, it is explicitly shown how the entanglement measure can be
calculated for the case of a resonant Bose-Einstein condensate with several
generated topological coherent modes. The resonant generation of these modes
and their temporal behaviour can be governed by external modulating fields.
Consequently, the evolutional features of entanglement can also be regulated.
This suggests new possibilities for quantum information processing and quantum
computing.

\vskip 5mm

{\bf Acknowledgement}

\vskip 3mm

I am very grateful for discussions to E.P. Yukalova. Financial support
from the Heisenberg-Landau Program is appreciated.


\newpage

\begin{thebibliography}{99}
\bibitem{1}
E. Schr\"odinger, {\it Proc. Camb. Phil. Soc.} {\bf 31}, 555 (1935).

\bibitem{2}
C.P. Williams and S.H. Clearwater, {\it Explorations in Quantum Computing}
(Springer, New York, 1998).

\bibitem{3}
M.A. Nielsen and I.L. Chuang, {\it Quantum Computation and Quantum Information}
(Cambridge University, New York, 2000).

\bibitem{4}
J.M. Raimond, M. Brune, and S. Haroche, {\it Rev. Mod. Phys.} {\bf 73}, 565
(2001).

\bibitem{5}
V. Vedral, {\it Rev. Mod. Phys.}, {\bf 74}, 197 (2002).

\bibitem{6}
A. Galindo and M.A. Martin-Delgado, {\it Rev. Mod. Phys.} {\bf 74}, 347 (2002).

\bibitem{7}
T.J. Havel {\it et al.}, {\it Am. J. Phys}. {\bf 70}, 345 (2002).

\bibitem{8}
M. Keyl, {\it Phys. Rep.} {\bf 369}, 431 (2002).

\bibitem{9}
C.H. Bennett, D.P. DiVincenzo, J.A. Smolin, and W.K. Wooters, {\it Phys.
Rev.} {\bf A 54}, 3824 (1996).

\bibitem{10}
V. Vedral and M.B. Plenio, {\it Phys. Rev.} {\bf A 57}, 1619 (1998).

\bibitem{11}
G. Boffetta, M. Cencini, M. Falconi, and A. Vulpiani, {\it Phys. Rep.} {\bf 356},
367 (2002).

\bibitem{12}
J. von Neumann, {\it Compos. Math.} {\bf 6}, 1 (1938).

\bibitem{13}
V.I. Yukalov, {\it Physica} {\bf A 310}, 413 (2002).

\bibitem{14}
A.J. Coleman and V.I. Yukalov, {\it Reduced Density Matrices} (Springer,
Berlin, 2000).

\bibitem{15}
A.S. Parkins and D.F. Walls, {\it Phys. Rep.} {\bf 303}, 1 (1998).

\bibitem{16}
F. Dalfovo, S. Giorgini, L.P. Pitaevskii, and S. Stringari, {\it Rev.
Mod. Phys.} {\bf 71}, 463 (1999).

\bibitem{17}
P.W. Courteille, V.S. Bagnato, and V.I. Yukalov, {\it Laser Phys.} {\bf 11} 659
(2001).

\bibitem{18}
V.I. Yukalov, {\it Statistical Green's Functions} (Queen's University, Kingston,
1998).

\bibitem{19}
M. Greiner et al., {\it Nature} {\bf 415}, 39 (2002).

\bibitem{20}
G. Kalosakas, K.\O . Rasmussen, and A.R. Bishop, {\it Phys. Rev. Lett.}
{\bf 89}, 030402 (2002).

\bibitem{21}
M.A. Andreata and V.V. Dodonov, {\it J. Phys.} {\bf A 35}, 8373 (2002).

\bibitem{22}
V.I. Yukalov, E.P. Yukalova, and V.S. Bagnato, {\it Phys. Rev.} {\bf A 56},
4845 (1997).

\bibitem{23}
V.I. Yukalov and E.P. Yukalova, {\it J. Phys.} {\bf A 35}, 8603 (2002).

\bibitem{24}
V.I. Yukalov, E.P. Yukalova, and V.S. Bagnato, {\it Phys. Rev.} {\bf A 66},
043602 (2002).

\bibitem{25}
V.I. Yukalov, E.P. Yukalova, and V.S. Bagnato, {\it Laser Phys.} {\bf 12},
231 (2002).
\end{thebibliography}
\end{document}